\documentclass[fleqn,usenatbib]{mnras}

\usepackage{newtxtext}
\usepackage{newtxmath}

\usepackage{natbib}
\usepackage[T1]{fontenc}
\usepackage{aecompl}

\usepackage{graphicx}							
\usepackage{bm}									
\graphicspath{{graphics/}}					
\usepackage{amsmath}							
\usepackage{amssymb}							
\usepackage{physics}							
\usepackage[nameinlink, capitalize]{cleveref}	
\usepackage{hyperref}							
\usepackage{array}								
\usepackage{xcolor}								

\newcommand{\erad}{\theta_\mathrm{E}}

\newcommand{\prob}[2]{\mathrm{Pr}\left(#1\vert#2\right)}
\newcommand{\prior}[1]{\mathrm{Pr}\left(#1\right)}

\newcommand{\Msun}{M_\odot}
\newcommand{\Nsub}{N_\mathrm{sub}}

\newcommand{\Mhm}{M_\mathrm{hm}}
\newcommand{\fsub}{f_\mathrm{sub}}
\newcommand{\Menc}{M_{2\erad}}

\newcommand{\musub}{\mu_\mathrm{sub}}
\newcommand{\mudet}{\mu_\mathrm{sub}}
\newcommand{\mcutoff}{m_\mathrm{cutoff}}
\newcommand{\mmin}{m_0}
\newcommand{\mmax}{m_1}
\newcommand{\mmap}{m_{\mathrm{map}}}
\newcommand{\Mtot}{M_\mathrm{tot,sub}^\mathrm{CDM}}
\newcommand{\pdet}{p_\mathrm{det}}
\newcommand{\fdet}{f_\mathrm{det}}

\newcommand{\mupop}{\mu_\mathrm{pop}}
\newcommand{\npop}{n_\mathrm{pop}}
\newcommand{\mpop}{m_\mathrm{pop}}
\newcommand{\spop}{\sigma_\mathrm{pop}}

\newcommand{\vmax}{v_\mathrm{max}}
\newcommand{\rmax}{r_\mathrm{max}}
\newcommand{\Mmax}{M_\mathrm{max}}

\newcommand{\modelone}{\mathcal{M}_1}
\newcommand{\modeltwo}{\mathcal{M}_2}
\newcommand{\modelthree}{\mathcal{M}_3}

\newcommand{\cneeded}{(\textcolor{cyan}{Author, \the\year{}})}

\defcitealias{ORiordan2024}{O'R24}
\defcitealias{ORiordan2023}{O'R23}

\title[Detecting the hidden population]{Detecting the hidden population of low-mass haloes in strong lenses}

\author[C. M. O'Riordan]{
	Conor M. O'Riordan$^{1}$\thanks{E-mail: conor@mpa-garching.mpg.de}
	\\
	$^{1}$Max Planck Institut f\"{u}r Astrophysik, Karl-Schwarzschild-Stra{\ss}e 1, 85748 Garching bei M{\"u}nchen, Germany\\
}
\date{Accepted XXX. Received YYY; in original form ZZZ}
\pubyear{\the\year{}}

\begin{document}

	\label{firstpage}
	\pagerange{\pageref{firstpage}--\pageref{lastpage}}
	\maketitle 

	\begin{abstract}
		A generic prediction of particle dark matter theories is that a large population of dark matter substructures should reside inside the host haloes of galaxies. In gravitational imaging, strong gravitational lens observations are used to detect individual objects from this population, if they are large enough to perturb the strongly lensed images. We show here that low-mass haloes, below the individually detectable mass limit, have a detectable effect on the lensed images when in large numbers, which is the case in cold dark matter (CDM). We find that, in CDM, this population causes an excess of 40 per cent in the number of detected subhaloes for HST-like strong lens observations. We propose a pseudo-mass function to describe this population, and fit for its parameters from the detection data. We find that it mostly consists of objects two orders of magnitude in mass below the detection limit of individual objects. We show that including this modification, so that the effect of the population is correctly predicted, can improve the available constraints on dark matter from strong lens observations. We repeat our experiments using models that contain varying amounts of angular structure in the lens galaxy. We find that these multipole perturbations are degenerate with the population signal. This further highlights the need for better understanding of the angular mass structure of lens galaxies, so that the maximum information can be extracted from strong lens observations for dark matter inference.

	\end{abstract}


	
	\section{Introduction}
\label{sec:introduction}
In particle dark matter theories, dark matter particles form overdensities called haloes which are steadily accreted by larger haloes. The largest haloes become hosts to galaxies and galaxy clusters, but many of the smaller haloes, or subhaloes, remain. The number density, mass, and mass-density profile of these subhaloes depend on the properties of the dark matter particle in ways that can be predicted by numerical simulations \citep{Springel2008,Vogelsberger2012}.

Galaxy-galaxy strong lensing observations provide a competitive probe of the properties of dark matter at these sub-galactic scales \citep[][for review]{Vegetti2023}. Subhaloes, or field haloes along the line of sight, have a gravitational lensing effect on the lensed images, secondary to that of the main lensing galaxy. If this effect can be reliably detected, information about the properties of dark matter can be inferred. Different methods have been proposed to do this.

In the gravitational imaging method, the lens galaxy is first modelled with some mass distribution which is smooth on the scale of the lens galaxy. Then, free-form, pixellated corrections to the lensing potential are found which improve the fit, with some regularisation condition to impose smoothness \citep{Vegetti2009}. The density corrections corresponding to the corrected potential are inspected. If a dark perturbing object exists in the field of view, it appears as a concentrated overdensity in the pixellated grid \citep{Vegetti2012,Cao2025}. Its properties can then be measured by fitting parametric models, comparing with the free-form corrected density map. The significant advantage of gravitational imaging is its agnosticism towards the specific form of the perturbation. No assumption needs to be made on the concentration of the perturbing object, its mass profile, or the ultimate number of perturbing objects in the field of view. If the data favour such concentrated overdensities, they will appear in the corrected convergence map without any necessary assumptions from the investigator on their properties.

To turn detections and non-detections from gravitational imaging into an inference on the dark matter model, a second, more costly, step is necessary. This is the computation of the so-called `sensitivity function'. This returns the smallest detectable mass of perturber in each pixel in the observation \citep[e.g.][]{Despali2022}. From this, the expected number of detectable objects given a dark matter model can be calculated, and the probability of that model calculated from the number of detections. In most cases with currently available data the expected number of detections, and actual detections, is zero, so computing the sensitivity map is essential for making use of all the available data.

Traditionally, the sensitivity function for an observation is computed by modelling a similar simulated image with and without a perturber of a given test mass, position and concentration. Bayesian model comparison between the models with and without perturbers gives the significance of a detection as a function of position and mass. Choosing a detection significance then gives the smallest detectable mass in each pixel. The computational cost is large and assumptions must be made about the perturber's concentration and redshift, or these variables are also mapped over, further increasing the cost. While the initial gravitational imaging step is highly flexible, data driven, and does not rely on assumptions about the perturber, the final step removes much of this flexibility by fixing the inference to whatever halo properties the sensitivity function is computed with.

To overcome this issue, especially in the context of large numbers of new strong lenses becoming available from e.g. Euclid, forward-modelling methods have been proposed \citep[e.g.][]{Brehmer2019,Gilman2020,AnauMontel2023,WagnerCarena2024,Filipp2024,Coogan2024}. These methods treat the specifics of individual detections essentially as nuisance parameters, and instead seek to infer the dark matter properties directly. Typically, strong lens data are compared with large amounts of simulated data, simulated according to different dark matter models with the entire predicted population of low-mass haloes injected into the image. The likelihood of each dark matter model is then directly computed via comparison with the data. These methods fall in the category often described as approximate Bayesian computation or simulation based inference.

An obvious advantage is that the dark matter model is directly inferred, without the need to explicitly and expensively calculate the sensitivity. But this inference can be relied upon only as far as the assumptions made in producing the simulated data. This is also true for methods which forward model individual perturbers without gravitational imaging. In some cases these have been in good agreement with gravitational imaging \citep{Nightingale2024}. Other cases have produced false positive results, demonstrating the inherent risk of such a method \citep{Hezaveh2016,Stacey2025}.

However, forward-modelling methods have another significant advantage in that they can account for the effect of the entire population of perturbers, not just those large enough to be individually detectable with gravitational imaging. After all, in CDM, we expect hundreds or thousands of subhaloes larger than $10^6\Msun$ to populate the haloes of lens galaxies. The sensitivity mapping necessary for dark matter inference using gravitational imaging does not currently account for this effect, if it exists. A correctly implemented forward-modelling method could detect or rule out such an effect, resulting in a more powerful inference on the dark matter model than considering individually detected objects only. Many of the works cited above include this effect implicitly, but do not show systematically that it exists, or measure its properties.

In this paper, we test for this effect systematically using mock strong lens images and a previously developed machine learning subhalo detection and sensitivity mapping method. We compare the size of the effect in different dark matter models and with different subhalo masses. We also test for a degeneracy between this effect and the angular structure in the lens galaxy.

The paper is organised as follows. In \cref{sec:method} we detail the method for detecting the effect of the population. In \cref{sec:results} we present our results, and in \cref{sec:discussion,sec:conclusions} we discuss possible issues and summarise the paper. Throughout we assume a Planck 2015 cosmology \citep{Planck2015}.
	\section{Method}
\label{sec:method}
\begin{figure}
    \includegraphics[width=1.0\columnwidth]{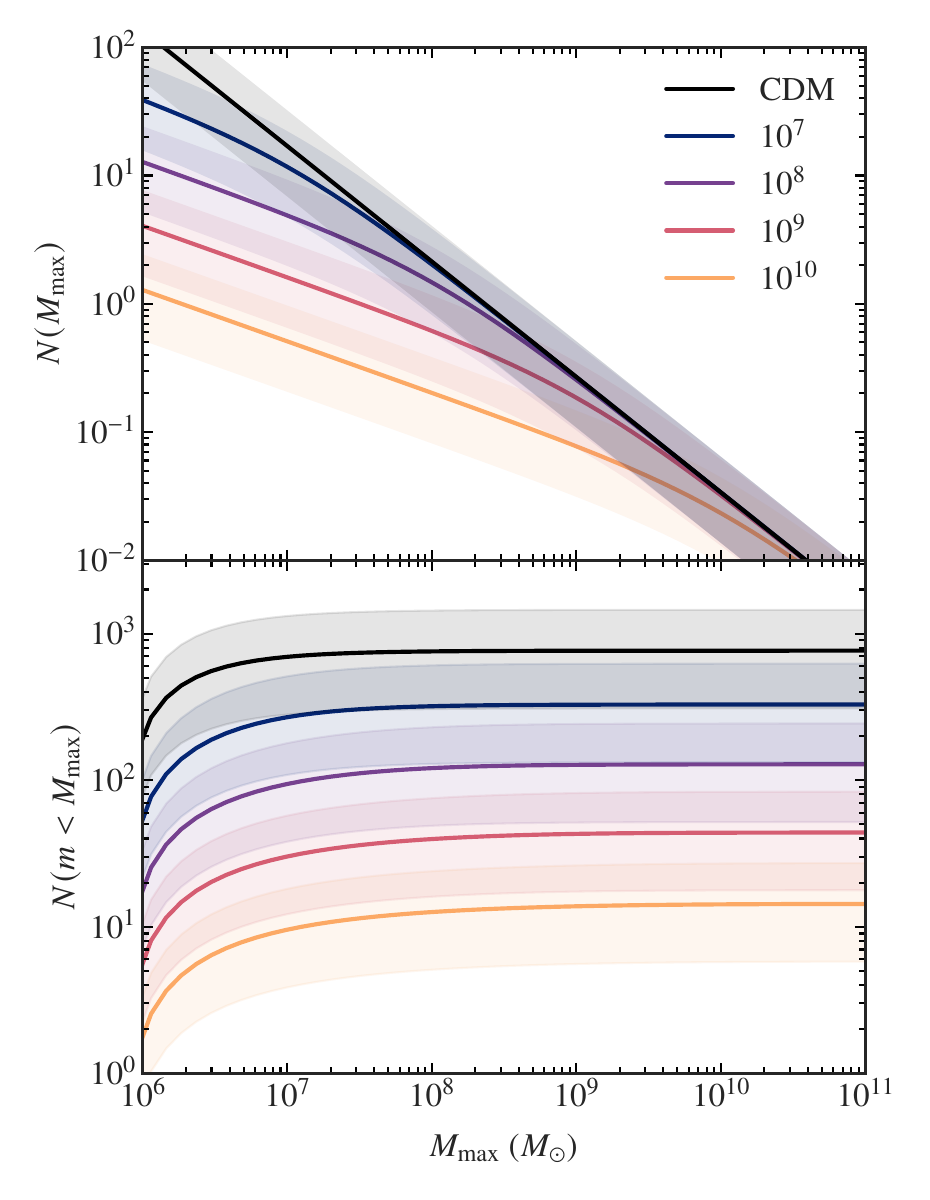}
    \caption{\label{fig:mass-function} Subhalo mass function for the lenses used in this work, with different values of $\Mhm$ indicated by colour. The lines give the median value and the shaded areas show the $1\sigma$ range over the $100$ lenses. Upper frame: the number of subhaloes per mass bin with bins of size 0.1 dex in $\Mmax$. Lower frame: The cumulative distribution going from left to right, i.e., the total number of subhaloes in a lens with mass below the $\Mmax$ $x$-axis value. For example, in CDM, a typical lens used in this work has between $300$ and $1100$ subhaloes within $2\erad$ in the chosen mass range, depending on the enclosed mass of the lens, $\Menc$.}
\end{figure}

We use the machine learning-based sensitivity mapping and substructure detection method introduced in \citet[][hereafter \citetalias{ORiordan2023}]{ORiordan2023} and extended in \citet[hereafter \citetalias{ORiordan2024}]{ORiordan2024}. Our results are based on applying these models to different realisations of the same $100$ lens systems used in \citetalias{ORiordan2024}, which were selected to resemble SLACS and BELLS Einstein ring systems previously studied for substructure. In this work, we only consider the effect of subhaloes, i.e., member haloes of the main lens galaxy halo, and exclude haloes along the line of sight.

\subsection{Subhalo mass function}

The number $n$ of subhaloes with mass $m$ in a unit area on the sky is proportional to the subhalo mass function
\begin{equation}                 
	\label{eq:mass-function}
	\derivative{n}{m} \propto m^{\alpha_1}\left[1+\left(\alpha_2\frac{\Mhm}{m}\right)^\beta\right]^{\gamma},
\end{equation}
where the constants have the following values: $\alpha_1=-1.9$, $\alpha_2=1.1$, $\beta=1.0$, and $\gamma=-0.5$. These are obtained from galaxy formation simulations in \citet{Lovell2020b}, re-fit to be in terms of $\Mmax$, which is the definition of subhalo mass used in this work. The parameter $\Mhm$ is the half-mode mass, which is the mass scale in a warm dark matter (WDM) model at which the power-spectrum is suppressed by half compared to that of CDM. It sets the scale below which a suppression in structure formation occurs, and is inversely proportional to the dark matter particle mass.

The expected number of substructures, $\musub$, in a lens with Einstein radius $\erad$, in a projected area $A$, between two mass limits, $\mmin$ and $\mmax$, is obtained by integrating \cref{eq:mass-function},
\begin{equation}
	\label{eq:mu-integral}
	\mu_{\mathrm{sub}}=\fsub^\mathrm{CDM}\frac{A}{4\pi\erad^2}\frac{\Menc}{\Mtot} \displaystyle\int_{\mmin}^{\mmax}\dv{n}{m}\dd{m},
\end{equation}
where the mass function is normalised such that $\fsub^\mathrm{CDM}$ is the fraction of mass in substructure within $2\erad$ in CDM, $\Menc$ is the mass of the lens enclosed in $2\erad$, and $A$ is the area on the sky. The definition of $\fsub^\mathrm{CDM}$ requires the term $\Mtot$ in the denominator, which is simply the total mass between the mass limits in a CDM mass function, given by
\begin{equation}
	\label{eq:norm-integral}
	\Mtot=\int_{\mmin}^{\mmax}m^{\alpha_1+1}\dd{m}.
\end{equation}
Normalising the mass function in this way ensures that WDM models, i.e., those with $\Mhm\gg0$, do not have a surplus of mass versus CDM in the range $m>\Mhm$. \Cref{fig:mass-function} shows the distribution of subhalo masses for different values of $\Mhm$ for the lenses used in this work, i.e., with the distribution of $\Menc$ in our chosen lenses, at a fixed $\fsub=10^{-2}$.

\subsection{Model training}
\label{sec:multipole-models}
To train the machine learning models we simulate strong lens observations at HST-like resolution and depth. The lens galaxies are power-law ellipsoids and the source surface brightness distributions are taken from a de-noised sample of Hubble Deep Field galaxy observations. The light profile of the main lens galaxy, simulated as a single S\'ersic profile, is subtracted up to its remaining Poisson noise. Different sources are used in training and testing to avoid overfitting.

To produce training data, lens and source model parameters are drawn from uniform distributions where ever possible .Training examples include either some number of substructures between 1 and 4, or no substructures, with equal probability. The subhaloes in training are NFW profiles with maximum circular velocity $\vmax$ drawn log-uniformly from the range $10 \leq \vmax/\mathrm{km\,s}^{-1}\leq 158$ and concentration, parametrised by the radius, $\rmax$, at which maximum velocity occurs, also drawn log-uniformly from the range $1\leq \rmax/\mathrm{kpc}\leq28$. In this way, the machine learning model does not `learn' any specific relationship between mass and concentration. These ranges give subhaloes which roughly have the mass range $10^7\leq\Mmax/\Msun\leq10^{11}$. The mass $\Mmax$ is the mass enclosed by $\rmax$ and is the definition of subhalo mass we use in the rest of the paper. We assume that the form of the density profile, i.e. an NFW, is the same across the entire mass range. In reality this is likely not the case due to the differing strength of baryonic effects at different mass scales \citep{Heinze2024}.

As in \citetalias{ORiordan2024}, we account for non-ellipticity in the lens mass distribution using multipole components. These are Fourier-like perturbations to the main lens density profile. To examine their effect on substructure detection, three versions of the model are trained. $\modelone$ includes no multipole perturbations in the training data. $\modeltwo$ includes them only up to amplitudes of $1$ per cent, and $\modelthree$ includes them up to 3 per cent. $\modeltwo$ and $\modelthree$ use multipoles of orders $1$, $3$, and $4$, with the amplitude of each order independently drawn from the model-dependent prior. $\modelone$ essentially represents most strong lensing models used until recently, a power-law ellipsoid plus external shear. $\modeltwo$ represents the best extension to $\modelone$, with non-elliptical components informed by the isophotes of elliptical galaxies, a model which is being used more commonly at present. $\modelthree$ represents an extreme extension of $\modeltwo$, so that effects due to multipoles can be more easily borne out.

 We then select a sample of $100$ lens systems from a catalogue of millions of simulated realistic strong lenses generated in \citetalias{ORiordan2023} according to the method of \citet{Collett2015}. We make the following cuts
 \begin{equation}
	\label{eq:conditions}
	\begin{split}
		&0.8\leq\erad/\mathrm{arcsec}\leq1.6,\\
		&\beta/\erad < 0.2,\\
		&q>0.5,\\
		&50<\mathrm{S/N}<300,
	\end{split}
\end{equation}
where $\erad$ is the Einstein radius, $\beta$ the projected radial position of the source galaxy, $q$ is the axis ratio of the lens galaxy mass profile, and S/N here is the total integrated signal to noise ratio of the lensed images only, computed according to \citet{ORiordan2019}, which is the definition of S/N which correctly predicts the uncertainty on the lens mass model parameters. We randomly choose $100$ lenses from the 1798 which satisfy the conditions. This selection intentionally returns systems most similar to the SLACS \citep{Bolton2006} and BELLS GALLERY \citep{Brownstein2012} samples used in previous dark matter studies. They are the same mock objects used in \citetalias{ORiordan2024} (their Fig. 1). These are effectively a subset of the parameter space of the training data, but have realistic distributions of Einstein radii, lens and source redshift, S/N, etc. The real HDF objects used as source galaxies in this sample, which is used for all of our results, were not part of the training set, i.e., the models have never `seen' these source galaxies before.

The machine learning models are all 50 layer ResNet architectures \citep{He2015}. They are trained to minimise the cross-entropy loss with $4\times10^6$ images, with a tenth that number used for computing validation loss. A parameter sweep is performed at the start to find the optimal learning rate. During training the learning rate is decayed by a factor $10^{-0.25}$ every time the validation loss does not improve for 10 epochs. After three consecutive decays with no validation loss improvement the model is assumed to have converged and training is stopped. Typically this takes between 500 and 1000 epochs. After training, we use the machine learning models in two ways: to create sensitivity maps for the chosen sample of $100$ lenses, and as substructure detectors in mock observations of those lenses containing different substructure populations.

\subsection{Sensitivity mapping}
To create a sensitivity map, the models are shown realisations of the same strong lens system with a test mass substructure that moves across all pixels and over a range of masses. Each realised image is passed through the model and the probability of the positive class, i.e., that some amount of substructure is present, is recorded. A probability threshold is chosen, in this case $5\sigma$. The sensitivity map itself then follows in a straightforward way, as simply the lowest mass of subhalo in each pixel where the detection probability reached the threshold. We name this mass $\mmap$. We compute a sensitivity map for each lens in this way using a test mass range of $10^7<\Mmax<10^{11}$, with 21 mass steps spaced log-uniformly. At each mass step, the $\rmax$ value of the subhalo is set to the median value from the $\vmax$-$\rmax$ relation of \citet{Moline2022}.

\begin{figure}
    \includegraphics[width=1.0\columnwidth]{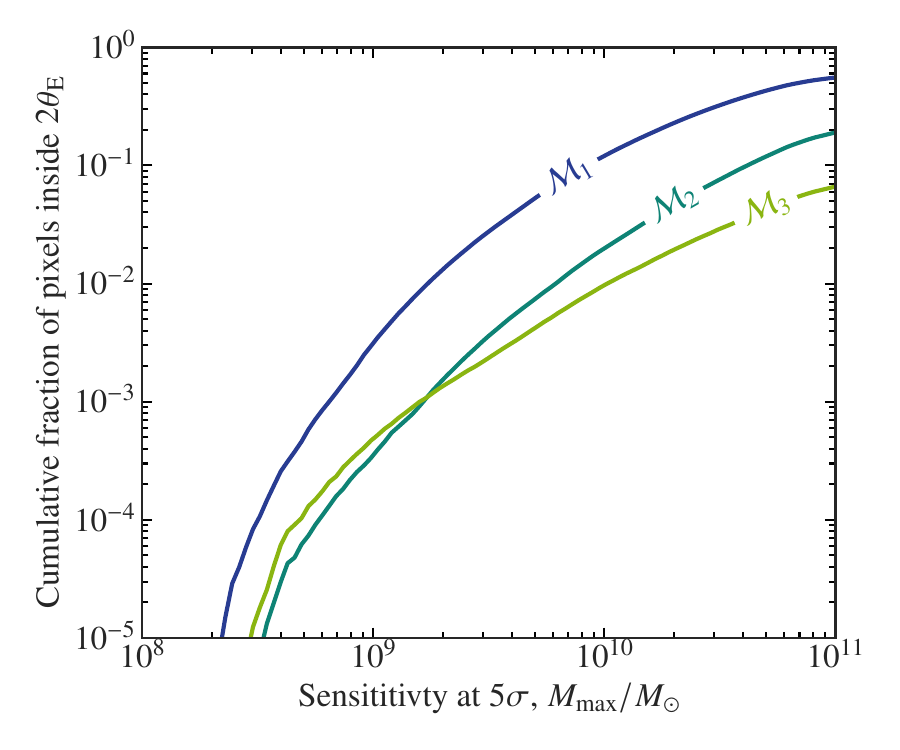}
    \caption{\label{fig:sensitivity} Cumulative fraction of sensitivities for pixels inside $2\erad$ for the $100$ mock lenses used in this work, for different multipole models.}
\end{figure}

\Cref{fig:sensitivity} shows the distributions of sensitivity across all pixels within $\erad$ in the $100$ mock lenses used in this work, for the three trained models. \Cref{table:sensitivity} includes some useful point statistics. As shown in \citetalias{ORiordan2024}, the addition of multipoles mostly reduces the sensitive area away from the lensed images, where only the largest subhaloes would be detectable. Sensitivity close to the lensed images, where the detectable mass is smallest, is less effected.

\begin{table}
    \centering
    \renewcommand{\arraystretch}{1.2}
    \begin{tabular}{p{0.5cm} p{1.4cm} p{1.2cm} p{0.9cm} p{0.9cm} p{0.9cm}}
        & & & \multicolumn{3}{c}{Fraction sensitive below}\\
        Model & Multipoles & Minimum sensitivity & $10^9\Msun$ $[\times 10^{-3}]$ & $10^{10}\Msun$ $[\times 10^{-2}]$ & $10^{11}\Msun$ $[\times 10^{-1}]$\\
        \hline
$\mathcal{M}_1$ & None & $10^{8.26}$ & $3.13$ & $12.02$ & $5.55$\\
$\mathcal{M}_2$ & $<1$ per cent & $10^{8.37}$ & $0.35$ & $1.98$ & $1.90$\\
$\mathcal{M}_3$ & $<3$ per cent & $10^{8.39}$ & $0.49$ & $0.97$ & $0.66$\\
        \hline
    \end{tabular}
    \caption{\label{table:sensitivity} Sensitivity statistics for the $100$ lenses used in this work, for the three models with different multipole strengths. The columns area: the name of the model, the allowed range of multipole amplitudes during training, the minimum sensitivity, and the fraction of pixels within $2\erad$, over all lenses, which have sensitivity below the indicated masses.}
\end{table}

\begin{figure}
    \includegraphics[width=1.0\columnwidth]{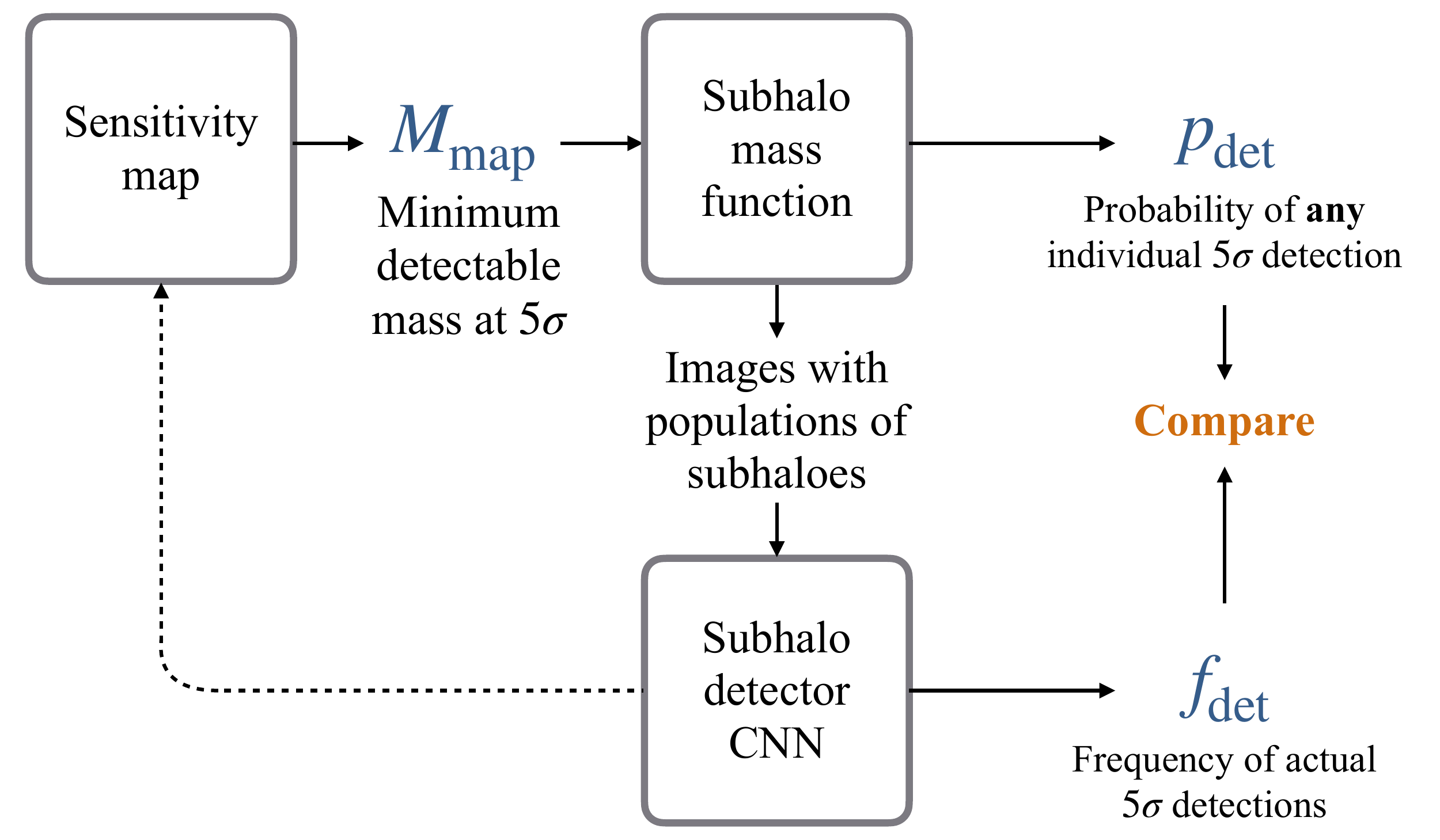}
    \caption{\label{fig:flowchart}Method to test the detectability of populations of individually undetectable substructures. A subhalo mass function, effectively the dark matter model, is chosen first, parametrised by $\Mhm$ and $\fsub$. From this populations of subhaloes are drawn to produce images. The images are run through a subhalo detector CNN, which is the same used to create the sensitivity map for each lens. The sensitivity map and the subhalo mass function give $\musub$, the mean number of detectable objects, and in turn $\pdet$, the probability that the CNN makes a detection. The actual number of detections made by the CNN is $\fdet$. Comparing these quantities is the basis for the results in this paper.}
\end{figure}

\subsection{Detecting a population of substructures}
To determine whether a population of objects below the detection threshold become detectable in larger numbers, we use the following procedure, which is also illustrated in \cref{fig:flowchart}. First, a dark matter model is chosen, parametrised by $\Mhm$. From this mass function and the sensitivity function computed as above, we find the expected number of detectable subhaloes in each lens system, $\mudet$. This is computed using \cref{eq:mu-integral} with $A$ as the pixel area, $\mmin=\mmap$, and $\mmax$ is the upper limit to the sensitivity calculation, in this case $\mmax=10^{11}\Msun$. Importantly, in the normalising term $\Mtot$, the lower limit is rather the lower limit on the substructure mass function, in this case $10^6\Msun$.

The number of detectable objects is Poisson distributed with mean $\musub$, so the probability that a given lens will have any detection, accounting for multiple individually detectable subhaloes, is
\begin{equation}
    \label{eq:pdet}
\pdet=1-\exp(-\musub),
\end{equation}
and for $\musub\ll1$, $\pdet\approx\musub$. The quantity $\pdet$ therefore describes the frequency with which we should expect a subhalo detection, if the sensitivity map correctly accounts for all of the detectable subhalo `signal' in the data, i.e., if the undetectable population of subhaloes has no effect.

Then for each lens system, we draw subhaloes from the same mass function in the range $10^6<\Mmax<10^{11}$. Their concentration is set by the $\vmax-\rmax$ relation of \citet{Moline2022} and used previously in \citetalias{ORiordan2023} and \citetalias{ORiordan2024}. They are positioned uniformly inside a circle of $2\erad$. We add the convergence of the subhaloes to that of the lens macro model and ray trace to the source plane to produce a simulated observation of that lens system with that realisation of the substructure population. We add realistic noise and PSF as in the training data of the models. We repeat this $10^4$ times, with different realisations of the population, for each of the $100$ lenses. We pass all these realisations through the machine learning models and record the frequency of detections, called $\fdet$. The measurement of $\fdet$ has the variance of a binomial distribution, i.e. $\sigma_{\fdet}^2=\pdet(1-\pdet)/N$, where $N=10^4$ is the number of trials.

Our interest is then in any difference between the prediction of $\pdet$ and the measurement of $\fdet$. If subhaloes are only detectable as individual objects which are large enough alone to perturb the lensed images to some measurable degree, then we should expect $\pdet\sim\fdet$. If however, objects at smaller, apparently undetectable, masses produce a detectable subhalo signal when present in large numbers, then we should observe an excess in $\fdet$ compared to $\pdet$.

	\section{results}
\label{sec:results}

\begin{figure*}
    \includegraphics[width=0.9\textwidth]{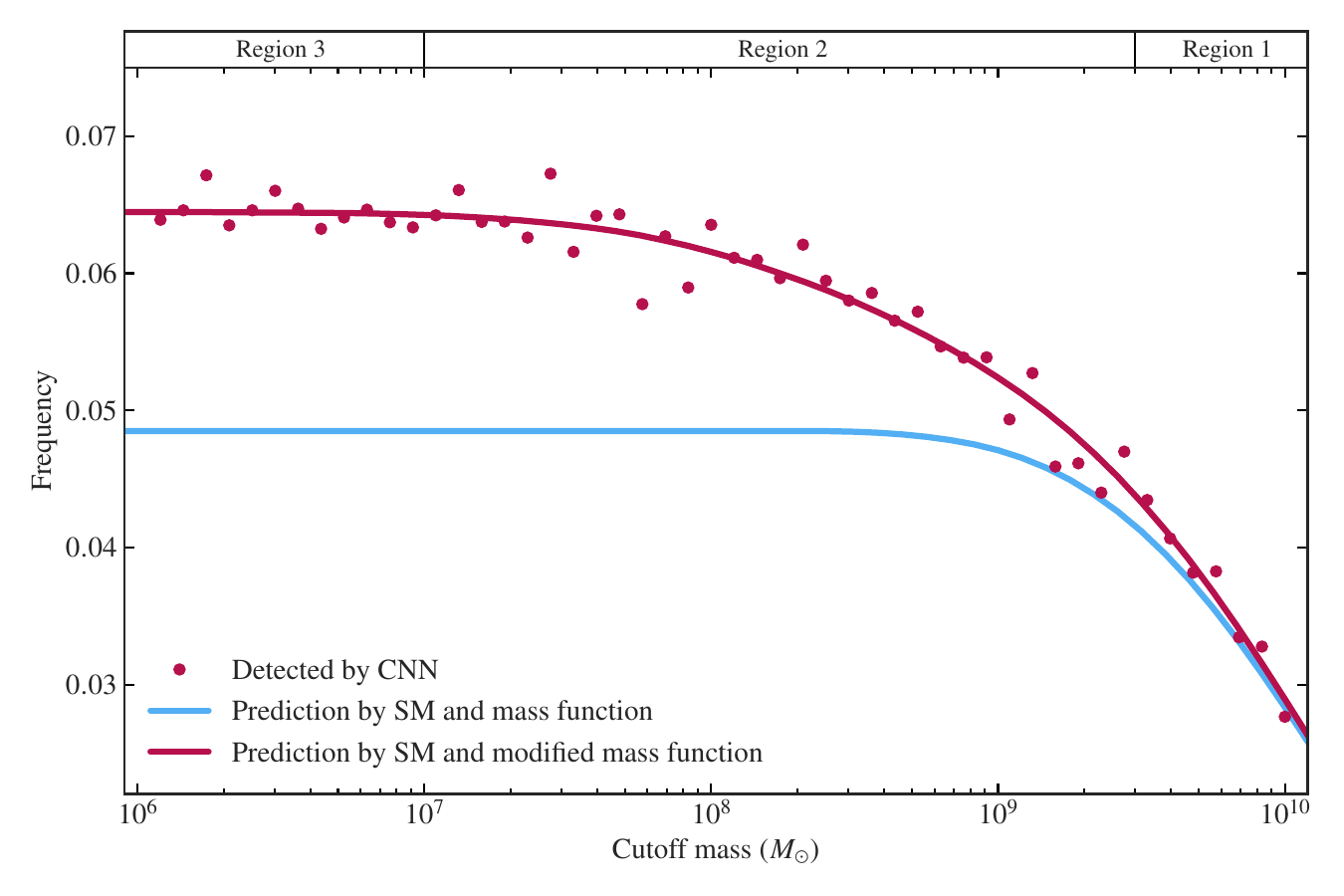}
    \caption{\label{fig:lower-bound-detections} Frequency of detections predicted by the sensitivity map, and achieved by the neural network as a function of the cutoff mass $\mcutoff$. For each realisation of each lens, $\mcutoff$ is drawn randomly and binned to produce this figure. The solid lines show the mean frequency per $\mcutoff$ bin and the shaded area the standard deviation within each bin. The standard deviation is large relative to the number of detections because the number of detections in general is small.}
\end{figure*}

\subsection{Excess detections in CDM and WDM}

We perform the comparison explained previously for mass functions with different values of $\Mhm$, using only the model trained without multipoles, $\modelone$. The effect of multipoles is examined in the next step. \Cref{table:results} shows the results for $12$ values of $\Mhm$. In all of this section, $\fsub$ is fixed with $\fsub=10^{-2}$.

In the CDM model, we observe a significant excess in the number of detected substructures compared to the value predicted by the sensitivity map. This implies that the sensitivity map does not capture all of the detectable subhalo signal, and that in fact, the population of low-mass haloes, themselves individually undetectable, are having a detectable effect on the lensed images. To quantify the excess in detections we fit a simple model to find the factor which would scale $\pdet$ to give the best agreement with $\fdet$, accounting for the uncertainty on $\fdet$. For the CDM case this excess is 39 per cent.

The size of this effect diminishes as the dark matter model becomes warmer, i.e., as $\Mhm$ increases. At $\Mhm=10^8\Msun$ the excess factor drops to  31 per cent and at $\Mhm=10^{10}\Msun$ to only 8 per cent. This reveals the population of low-mass haloes present in CDM but suppressed in WDM to be the source of the excess. Between $\Mhm=10^8\Msun$, and $\Mhm=10^{11}\Msun$ the mass suppression of the dark matter model crosses the sensitivity boundary of the data (compare with \cref{fig:sensitivity}). As these nominally undetectable lower mass haloes are suppressed in number, so too is the excess in detections. The non-zero excess in the very warmest models is due to a small number of low-mass haloes still remaining. This is because the mass function used here and plotted in \cref{fig:mass-function} does not have a sharp break at the half-mode mass.

\begin{table}
    \centering
    \begin{tabular}{l l l l l l}
        $\Mhm$ & $N_\mathrm{sub}$ & $\Mmax<\Mhm$ & $\hat{N}_\mathrm{d}$ & $N_\mathrm{d}$ & Excess\\
        $[\Msun]$ & & [per cent] & & &\\
        \hline
        CDM & $746.0$ & $0.0$ & $4.5\pm2.0$ & $6.54$ & $1.39$\\
        $10^{6.0}$ & $615.6$ & $0.0$ & $4.5\pm2.0$ & $6.54$ & $1.39$\\
        $10^{6.5}$ & $489.6$ & $55.4$ & $4.5\pm2.0$ & $6.43$ & $1.36$\\
        $10^{7.0}$ & $347.0$ & $77.6$ & $4.5\pm2.0$ & $6.41$ & $1.37$\\
        $10^{7.5}$ & $225.3$ & $87.8$ & $4.5\pm2.0$ & $6.27$ & $1.34$\\
        $10^{8.0}$ & $138.6$ & $93.0$ & $4.5\pm2.0$ & $6.08$ & $1.31$\\
        $10^{8.5}$ & $82.3$ & $95.8$ & $4.4\pm2.0$ & $5.68$ & $1.24$\\
        $10^{9.0}$ & $47.8$ & $97.5$ & $4.2\pm1.9$ & $5.23$ & $1.20$\\
        $10^{9.5}$ & $27.3$ & $98.5$ & $3.9\pm1.9$ & $4.52$ & $1.14$\\
        $10^{10.0}$ & $15.6$ & $99.2$ & $3.3\pm1.7$ & $3.59$ & $1.08$\\
        $10^{10.5}$ & $8.9$ & $99.6$ & $2.5\pm1.5$ & $2.63$ & $1.05$\\
        $10^{11.0}$ & $5.0$ & $100.0$ & $1.7\pm1.3$ & $1.72$ & $1.02$\\
        \hline
    \end{tabular}
    \caption{\label{table:results}Subhalo population and detection statistics for 12 tested values of $\Mhm$. The columns are: the half-mode mass, $\Mhm$; the average number of subhaloes per lens, $\Nsub$, in the mass range $10^6\Msun-10^{11}\Msun$; the percentage of that number which have a mass $\Mmax$ below the half-mode mass; the expected number of detections, $\hat{N}_\mathrm{d}$, in the sample of $100$ lenses, with associated binomial uncertainty; the average number of detections $N_\mathrm{d}$ over $10^4$ trials; and the excess factor. The excess is the result of fitting to the data and takes the uncertainty into account, and so is not straightforwardly $N_\mathrm{d}/\hat{N}_\mathrm{d}$.}
\end{table}

\subsection{Excess as a function of subhalo mass}

To explicitly find the source of the excess detections, we repeat the experiment with an artificially truncated mass function. Subhaloes are drawn as normal from a CDM mass function, but all subhaloes below a given cutoff mass, $\mcutoff$ are discarded before ray-tracing and passing the realisations through the neural networks to record detections. The cutoff mass is drawn log-uniformly for each realisation from the range $10^6<\mcutoff/\Msun<10^{11}$. The rate of detections predicted by the sensitivity map and mass function, $\pdet$ is computed as in the previous section, except the lower limit on the mass function integral, the value of the sensitivity map, is replaced with the cutoff mass if the cutoff mass is larger.

\Cref{fig:lower-bound-detections} shows the frequency of detections predicted by the sensitivity map, and made by the neural network as a function of $\mcutoff$\footnote{In the limiting case of $\mcutoff\rightarrow10^{6}\Msun$, the test is the same as the CDM case in the previous section, and we recover the same rate of both predicted and actual detections as in the first row of \cref{table:results}.}. The figure reveals three different regions of behaviour which we label at the top of the frame, from right to left. In region~$(1)$, for $\mcutoff$ of $10^{9.5}\Msun$ and above, all subhaloes are above the sensitivity limit of the data, and so the sensitivity map reliably predicts the frequency of detections. As $\mcutoff$ decreases, the predicted and detected frequencies both steadily increase, as more detectable subhaloes are added to the data.

In region~$(2)$, at $\mcutoff\sim10^{9.5}\Msun$, the two curves begin to diverge. \Cref{fig:sensitivity} shows that most pixels have no sensitivity below this mass. Accordingly, the predicted frequency of detections from the sensitivity map and mass function becomes flat as all newly added subhaloes are too small to be individually detected. However, the actual frequency of detections by the CNN increases until $\mcutoff\sim10^{7}\Msun$. In regime $(3)$, below $\mcutoff\sim10^{7}\Msun$, the excess of detections stops increasing. This shows that subhaloes in the intermediate mass range, i.e., in region $(2)$, have a detectable effect on the lensed images. Subhaloes at the lowest mass, i.e., for $\mcutoff\lesssim10^{7}\Msun$ and below, do not have a detectable effect.

\begin{figure}
    \includegraphics[width=1.0\columnwidth]{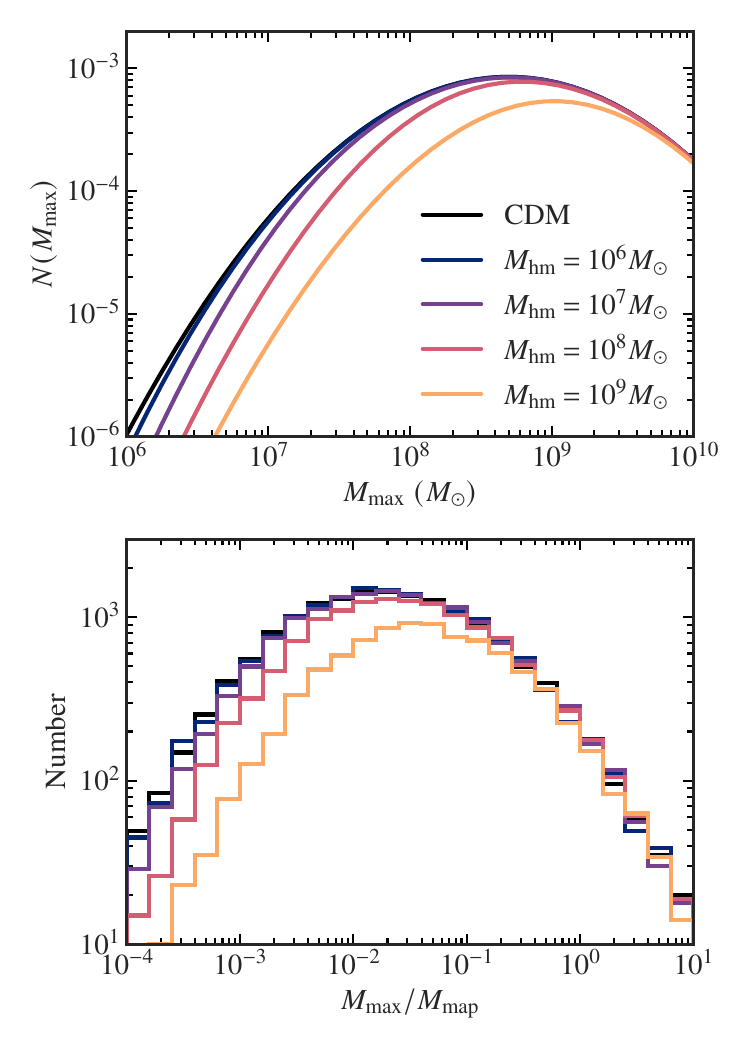}
    \caption{\label{fig:population-mass-function} Properties of the subhaloes in the detectable population for different values of $\Mhm$. Upper frame: the modified mass function in \cref{eq:mupop-integral} using the best fit values for our HST strong lenses. Lower frame: the ratio of the individual masses of the detectable population objects to the pixel sensitivities in our data, randomly drawn.}
\end{figure}

\subsection{Accounting for the detectable population}

To properly account for the extra detections due to the population of low-mass haloes we require a modification to the calculation of $\pdet$ such that haloes in the intermediate mass range increase the number of expected detections. We rewrite \cref{eq:pdet} to include a new term, $\mupop$,
\begin{equation}
    \label{eq:pdet-modified}
\pdet=1-\exp\left[-(\musub+\mupop)\right].
\end{equation}
The number $\mupop$ can be thought of as the number of extra detectable objects if all detectable `populations' were converted to equivalent individual objects. We calculate $\mupop$ using a modified subhalo mass function which artificially increases the number of objects of mass $m$ by a factor $g(m)$, that is,
\begin{equation}
	\label{eq:mupop-integral}
	\mupop=\fsub^\mathrm{CDM}\frac{\Menc}{\Mtot} \displaystyle\int_{\mmin}^{\mmax}\dv{n}{m}g(m)\dd{m},
\end{equation}
Note that this quantity is computed on a per lens basis, rather than per pixel, so there are no factors related to the area.

From \cref{fig:lower-bound-detections}, we know that $g(m)$ must have some functional form that goes to zero at both extremes of mass, but must peak in the intermediate mass range. A Gaussian is therefore a natural choice and we write $g(m)$ as,
\begin{equation}
    g(m)=n_\mathrm{pop}\exp\left[-\left(\frac{\log(m)-\log(m_\mathrm{pop})}{\sigma_\mathrm{pop}}\right)^2\right].
\end{equation}
This introduces three new parameters. The normalisation $n_\mathrm{pop}$ represents the strength of the effect in causing detections, and may be roughly interpreted as the ratio of the number of subhaloes in a detectable population to an equivalent individually detectable object. The mean and variance of $g(m)$ are $\mpop$ and $\spop^2$, but these are only weak indicators of the mass range of the detectable population. This is because of the shape of the mass function, with which $g(m)$ will be multiplied, where every decade lower in mass brings a decade more in number (see again \cref{fig:mass-function}). For example, if we find $\mpop=10^X\Msun$ with $\spop=1$ (one order of magnitude in mass), much more than half of the population members will have mass $\Mmax<10^X$.

We obtain the values of the three new parameters by MCMC fitting. At a given cutoff mass, we sample $n_\mathrm{pop}$, $m_\mathrm{pop}$ and $\spop$ and compute $\pdet$ as defined in \cref{eq:pdet-modified}. We calculate a log-likelihood that depends on the $\chi^2$ between the $\pdet$ prediction and the measured rate of detections, i.e., the points in \cref{fig:lower-bound-detections}. We assume the scatter on the points to be indicative of the uncertainty, as the statistical uncertainty is much smaller. We find the values
\begin{align*}
    \npop &= 4.73\,^{+1.40}_{-0.79}\times10^{-3},\\
    \mpop &= 10\,^{9.82\,^{+0.38}_{-0.28}} \Msun,\\
    \spop &= 1.04\,^{+0.17}_{-0.15},\\
\end{align*}
which produce the `modified' curve in \cref{fig:lower-bound-detections}.

Using the recovered values and \cref{eq:mupop-integral}, we plot the pseudo-mass function of this population in the upper frame of \cref{fig:population-mass-function}. Seeing as this population has a strong dependence on $\Mhm$, detecting its presence (or absence) in strong lens observations will improve constraints on the dark matter model. 

In the lower frame of \cref{fig:population-mass-function} we illustrate the size of the population members compared to typical detectable individual subhaloes. We draw subhaloes from the pseudo-mass function used in \cref{eq:mupop-integral} only and compare their masses with the sensitivity in a randomly drawn pixel from the $100$ mock lenses. For all values of $\Mhm$, the distribution peaks at $\sim2$ orders of magnitude below the individually detectable mass. The absolute masses of this population (upper frame) depend on the data, so the peak in the detectable population mass function between $10^8 \Msun$ and $10^9 \Msun$ is specific to the BELLS and SLACS-like HST observations used here. However, it is very likely that the relative masses of the detectable population (lower frame) do not change with angular resolution. This is because sensitivity scales straightforwardly with angular resolution \citep{Despali2022}. Extrapolating the results here to higher angular resolutions would suggest that strong lens observations with e.g. VLBI, where objects as small as $10^6\Msun$ are detectable, could be sensitive to populations of subhaloes whose constituent masses are $10^5\Msun$ or lower. Replicating the results in this paper at that angular resolution is currently computationally intractable.

The use of $\mupop$ here and its associated parameters is to correct the prediction for the rate of detections by the sensitivity map and the subhalo mass function, and to show that the observed excess in the rate of detections can be directly attributed to a low-mass halo population. It is useful for estimating the properties of the detectable population in this specific case where our machine learning models return a posterior probability only on the binary presence or absence of substructure. In a fully-forward modelling method \citep[e.g.][]{Brehmer2019} such a correction is not necessary, as the posterior on the dark matter model is already marginalised over all subhalo properties.

\subsection{Improved constraints on dark matter}

Here we consider the strength of constraint on the dark matter model, parametrised only by $\Mhm$, when the effect of the population either is or is not accounted for. We consider two simple instructive examples. In the first case, zero detections are made in our set of 100 lenses, which we label $\mathcal{D}_0$, and in the second case, 10 detections in 100 lenses, which we label $\mathcal{D}_{10}$. The case of 10 detections in 100 lenses represents an order of magnitude estimate of a realistic scenario given current substructure search efforts and the quality of the available data.

For each lens, the only data we use is whether an object is detected or not, in this case at a $5\sigma$ detection threshold. This is because it is the only information our CNN returns for each test image. We do not use any information about the number of detected objects in an individual lens, or the mass, concentration, or redshift of the objects. This means the constraint we obtain is extremely conservative, and a similar effort which took account of the properties of the detected objects would achieve stronger constraints.

The probability of a dark matter model described by $\Mhm$ and $\fsub$, given some dataset of detections and non-detections $\mathcal{D}$ is
\begin{equation}
    \label{eq:prob-non-detection}
    \prob{\Mhm}{\mathcal{D}} \propto \prior{\Mhm}\prior{\fsub}\prod_{i}^{\mathrm{ND}}\mathrm{e}^{-\mu_i}\prod_{j}^{\mathrm{D}}\mathrm{e}^{\,\mu_j},
\end{equation}
where $\prior{\Mhm}$ and $\prior{\fsub}$ are log-uniform priors on $\Mhm$ and $\fsub$, $\mu$ is the number of expected detections, itself a function of $\Mhm$ and $\fsub$, $\mathrm{ND}$ are the set of lenses with non-detections, and $\mathrm{D}$ is the set of lenses with detections. In \cref{eq:prob-non-detection}, $\mu_i$ can be calculated for the $i$th lens either accounting for the effect of the population, in which case $\mu_i=\musub+\mupop$, or without it, in which case it is simply $\mu_i=\musub$. To keep the comparison simple, the number of detections is kept the same whether the population is used or not, even though in reality, in the scenario where the population is present, there would also be $\sim 1.4$ times more detections.
\begin{figure}
    \includegraphics[width=1.0\columnwidth]{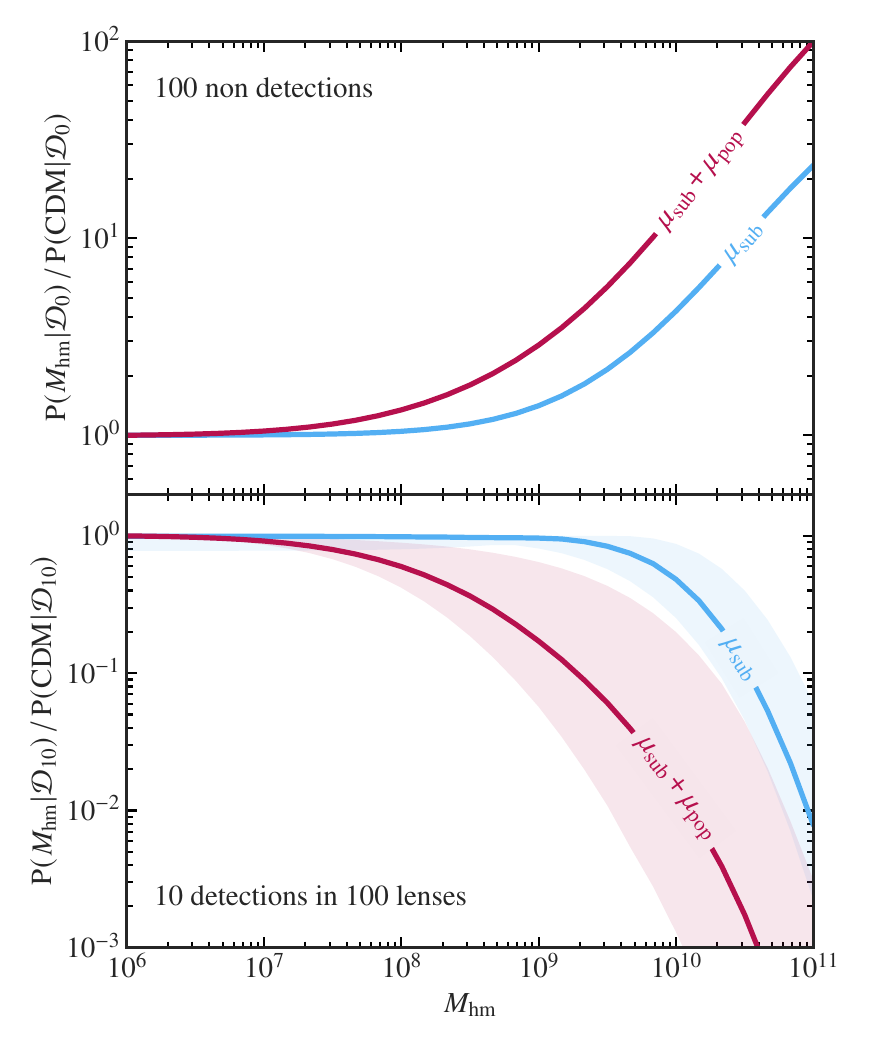}
    \caption{\label{fig:relative-constraints} The posterior probability of $\Mhm$, relative to that of CDM, i.e., $\Mhm\rightarrow0$, for two example scenarios. Upper frame: $\mathcal{D}_{0}$ (100 non-detections in HST strong lenses). Lower frame: $\mathcal{D}_{10}$ (10 detections in 100 lenses). The posterior can be calculated in two ways: including the population so that $\mu=\musub+\mupop$, or considering only $\musub$. In the $\mathcal{D}_{10}$ case, the exact constraint depends on in which lenses the detections are made due to their differing sensitivities. The shaded regions there represent the two extreme scenarios, with the solid line as the median constraint.}
\end{figure}

In \cref{fig:relative-constraints}, we plot the posterior on $\Mhm$ relative to CDM for both cases with a fixed $\fsub=10^{-2}$. For the $\mathcal{D}_{10}$ case, the result depends on exactly which lenses the detections and non-detections are in, as the difference in sensitivity can be large between lenses. We therefore sample many combinations and plot a range of possible constraints. The two curves show that accounting for the effect of a detectable population of low-mass haloes in our inference greatly improves our ability to constrain the dark matter model with the same data. In the example scenario of 100 non-detections of substructures in HST strong lenses, a rather extreme $\Mhm=10^{10}\Msun$ would only be favoured by 4:1 over CDM if the effect of the population is neglected. By including it, the same value is favoured over CDM at 14:1. At lower values of $\Mhm$ which are as yet not ruled out by astrophysical data, i.e., for $\Mhm<10^8\Msun$, the $\musub$ only inference would produce a result consistent with CDM. 

In the scenario where some lenses have detections, the constraints on $\Mhm$ are also improved by accounting for the population. In the median constraining case, CDM is only favoured over $\Mhm=10^{10}\Msun$ with odds of 2:1 when detections of a population are not included. When they are included, CDM is preferred over the same value with odds of $68$:$1$.

In both detection scenarios, CDM can be weakly distinguished from WDM all the way down to $\Mhm\sim10^7\Msun$. This is far below the traditionally considered sensitivity limit of HST data, suggesting that properly accounting for the effect of a population of low-mass substructures would allow for stronger constraints on dark matter models with already existing observations.

It is important to note that in this case the only information used to constrain the dark matter model is the binary presence of detectable substructure in each lens because this is the only data point that our neural network substructure detectors returns. In a full gravitational imaging study, more information about the number, mass, concentration, and redshift of the detected perturbers can be used to further improve the constraints.

\begin{table}
    \def\arraystretch{1.5}
    \centering
    \begin{tabular}{l l l l l}
        & Multipoles & $\npop$ & $\mpop$ & $\spop$ \\
        & & $[\times 10^{-3}]$ & $[\log(M/\Msun)]$ & \\
        \hline
        $\modelone$   & None & $4.73\,^{+1.40}_{-0.79}$ & $\,\,\,9.82\,^{+0.38}_{-0.28}$  & $1.04\,^{+0.17}_{-0.15}$\\
        $\modeltwo$   & $<1$ per cent & $3.33\,^{+3.89}_{-2.26}$ & $10.26\,^{+0.52}_{-0.70}$ &       $0.73\,^{+0.42}_{-0.43}$\\
        $\modelthree$ & $<3$ per cent & $0.64\,^{+2.80}_{-4.60}$ & $10.41\,^{+0.45}_{-3.60}$ &       $0.31\,^{+0.43}_{-0.16}$\\
        \hline
    \end{tabular}
    \caption{\label{tab:multipoles}Values for the modified mass function parameters found for three models using different amplitudes of multipole perturbation in their training data. Values are the median and the uncertainties are the 16th and 84th percentiles.}
\end{table}

\begin{figure}
    \includegraphics[width=1.0\columnwidth]{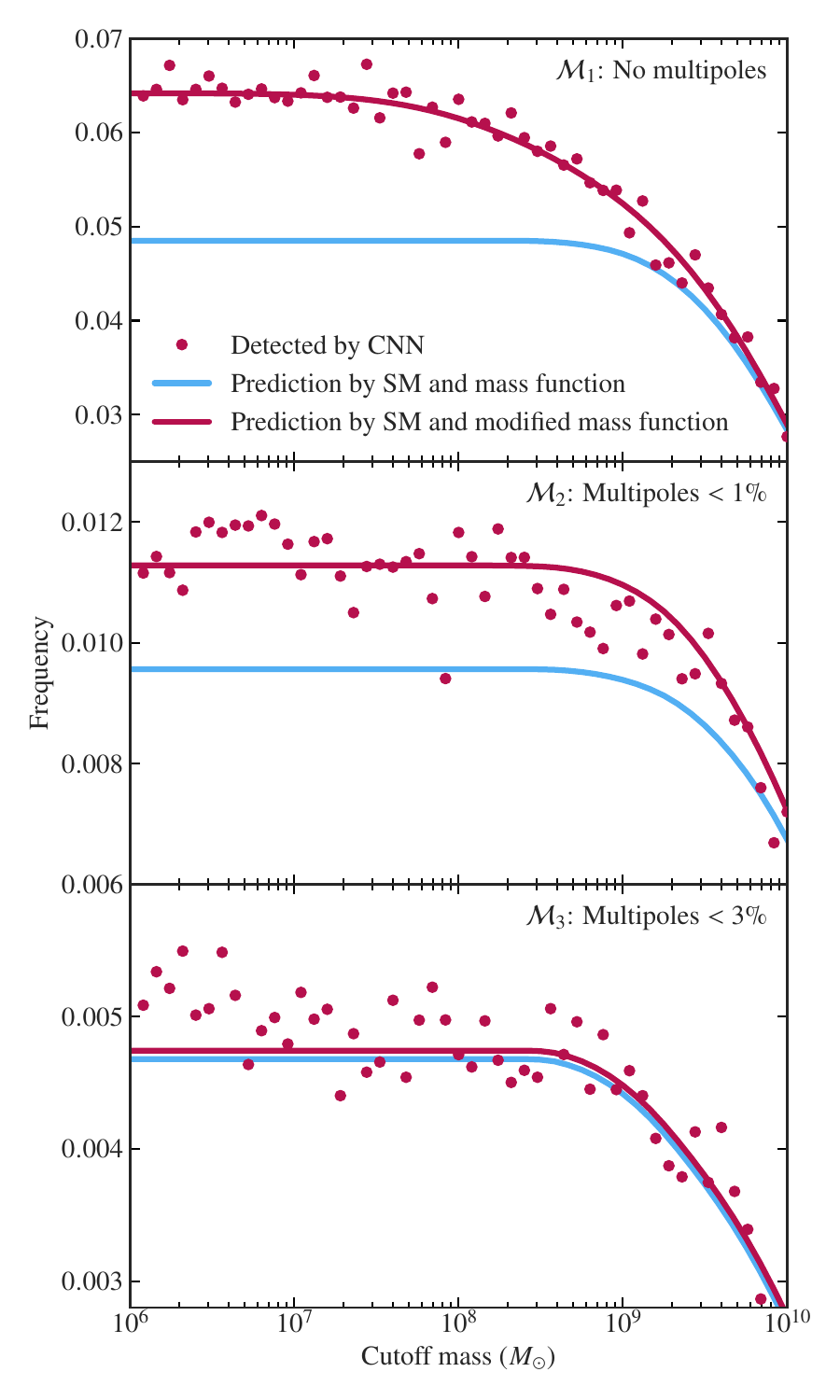}
    \caption{\label{fig:multipoles} The frequency of detections made by three different CNN models, compared with the frequency predicted by the mass function and sensitivity map. The three models differ in the strength of multipole perturbations allowed in the training data, labelled in each frame. The `modified' curves use the values in \cref{tab:multipoles}. The upper frame is identical to \cref{fig:lower-bound-detections}. Note especially the drastically different scales between the three frames.} 
\end{figure}

\subsection{Angular structure in the lens galaxy}
\label{sec:multipoles-results}
In \citetalias{ORiordan2024}, a degeneracy was identified between dark matter substructure and angular structure in the lens galaxy. This degeneracy is such that the number of detectable substructures in a strong lens is suppressed when the mass model for the lensing galaxy is allowed to contain angular structure beyond simple ellipticity. This angular structure is modelled using multipole perturbations, motivated by the observed isophotes of elliptical galaxies, which have perturbations beyond perfect ellipticity of order $\sim1$ per cent.

Using the same models from \citetalias{ORiordan2024} we repeat the experiment shown in \cref{fig:lower-bound-detections}. This time, we use the two models described in \cref{sec:multipole-models}. These models were trained with multipole perturbations in the lensing galaxy. This makes their substructure detections, and sensitivity maps, much more conservative. 

\Cref{fig:multipoles} shows the results, with the result for the no-multipole case, $\modelone$, included for comparison. \Cref{tab:multipoles} shows the values for the mass function parameters found from fitting to those data. The effect of the multipole perturbations is to suppress the number of detections overall, as already shown in \citetalias{ORiordan2024}, but to completely suppress the effect of the low-mass population. If $\npop$ is an accurate description of the relative strength of the effect, then it decreases by $\sim 30$ per cent for multipoles up to $1$ per cent, and up to $\sim 80$ per cent for multipoles up to $3$ per cent. In fact, MCMC fitting shows the posterior on $\npop$, and thus the effect itself, to be consistent with zero for the $3$ per cent case. It can be concluded that any effect from a population of very low-mass haloes, below the sensitivity limit, must be degenerate with the angular structure of the lens galaxy to a much greater extent even than for individually detectable perturbers.
	\section{Discussion}
\label{sec:discussion}

Here we discuss some limitations and implications of our work. We do not discuss aspects related to machine learning specific issues, e.g. quality of the training data, model architecture, etc. These issues were discussed for the same models used here in \citetalias{ORiordan2023} and \citetalias{ORiordan2024}. In \citetalias{ORiordan2024} a detailed comparison was made between the sensitivity maps produced by our machine learning models, and classical sensitivity maps produced by \citet{Despali2022}. We found the overall structure of the sensitivity maps to be the same. That is, for a given lens and source configuration, the topology of the sensitivity maps are similar. However, we find in general that the machine learning method produces more conservative sensitivity maps. This is for two reasons. First, its much larger `prior', i.e., its training data, includes all possible source and lens configurations, rather than the combinations which closely resemble the lens being modelled, as in the classical method. Second, and more importantly, is the extreme difficulty of the problem as presented to it, where almost all of the subhaloes presented in the training data and labelled as positive cases, are not detectable because they are too small or too far from the lensed images, or not concentrated enough. The result is an extremely cautious model where a $10\sigma$ detection in the classical method may be more like a $5\sigma$ detection in our case. In the context of this work, this conservative approach is actually an advantage, as it means the machine learning models never make spurious false positive detections unless a genuinely detectable subhalo signal is present in the data.

\subsection{Systematic uncertainty due to lens galaxy properties}
We find that the addition of large amplitude, circular multipoles to the lens galaxy model greatly reduces the detectability of the population of low-mass haloes. For the most extreme model, with multipoles $<3$ per cent, the effect is completely destroyed. In \citetalias{ORiordan2024} we trained models $\modeltwo$ and $\modelthree$, also used here, with circular multipoles. The deflection angle due to the circular multipoles is trivial to calculate, but this formulation adds perturbations in a circular basis to a mass distribution in an elliptical basis, and so for lenses which are far from circular, the mass distribution can become unrealistic.

Since these models were trained a computationally friendly calculation of the deflection angle in the elliptical basis has been found \citep{Paugnat2025}. In this basis, the angular perturbations are less localised and more closely follow the macro structure of the main lens galaxy. As \citet{Paugnat2025} show, elliptical multipoles produce smaller flux ratio anomalies in simulated lensed quasars than circular multipoles \citep{Cohen2024}. It should then be expected that the elliptical multipole perturbations are less degenerate with an individual low-mass halo, although this has not yet been systematically tested. 

The results in \cref{sec:multipoles-results} and the strong dependence on the angular structure allowed in the lens galaxy, however formulated, again underline the need for a comprehensive study of the angular mass-density structure in real lens galaxies. Although the use of elliptical multipoles is well motivated by the isophotes of elliptical galaxies, it is yet unclear whether the isophotes are an indicator of the mass distribution to the accuracy required for dark matter substructure studies. Therefore, the results in \cref{fig:multipoles} and \cref{sec:multipoles-results} can be thought of as presenting the worst-case scenario for low-mass halo population detection. This experiment would ideally be repeated using elliptical multipoles and a prior informed by measured multipoles in the total mass profiles of real strong lenses, or at least in simulated galaxies. Such measurements are currently too few in number to build an accurate prior. In such a test, it is likely that the result would lie somewhere between the $\modelone$ (\cref{fig:lower-bound-detections}) and $\modeltwo$ cases, presenting more optimistic prospects for constraining the low-mass halo population.

As well as the lens galaxy's angular structure, its emission can also interfere with the detection of dark substructures if not properly dealt with \citep[e.g.][]{Nightingale2024}. In this work, we assumed that the lens light is perfectly modelled and subtracted before the data are analysed for substructures. This is realistic for radio observations, where the lens galaxy typically has no emission \citep[e.g.][]{Powell2025} but for optical and NIR wavelengths, the lens galaxy light presents an issue. When detecting a population of substructures, whose effect is more diffuse than that of a single perturber, the degeneracy with the lens galaxy light is likely to be stronger.

\subsection{Prospects for dark matter inference}

We have shown that strong lens observations are sensitive to perturbations from populations of individually undetectable subhaloes. This population exists in a mass range which carries a large amount of information about the dark matter model, information which is presently discarded by the sensitivity map. In CDM, this population of objects with masses $\Mmax>10^6\Msun$ exists in abundance in every strong lens galaxy, so that studies which find an excess of detections in many lenses would obtain strong evidence in favour of CDM. In WDM models, this population is suppressed for masses $\Mmax<\Mhm$. Finding no excess of detections in many lenses would be evidence in favour of a value of $\Mhm$ above the lower mass limit of the population. In \cref{fig:population-mass-function} we show this to be up two orders of magnitude below the individual sensitivity limit. Although the constraints in \cref{fig:relative-constraints} are conservative, as they do not make use of mass and concentration information, taking account of the population effect would significantly improve the state of the art constraints on dark matter from strong lensing \citep{Enzi2021}. Scatter in the subhalo mass-concentration relation also means that CDM models produce some subhaloes with higher than average concentration. This further increases the observable difference between CDM and WDM, where these outliers are absent \citep{Amorisco2022}. In this work we required the properties of the simulated subhalo populations and the subhaloes used in the sensitivity map to correspond exactly for the same mass. As such, we could not account for the effect of this scatter on the dark matter constraints.

In self-interacting dark matter (SIDM), subhaloes can undergo a core-collapse process that increases their concentration to high values unobtainable in CDM \citep{Shinichiro2025}. This makes the subhaloes extremely efficient lenses and more easily detectable than their CDM counterparts. The mass range in which this core-collapse is most likely to occur depends on the interaction cross section in the SIDM model, which can be velocity (mass scale) dependent. If an individual core-collapsed subhalo becomes more detectable than its CDM counterpart due to increased concentration, it may also be the case that a population of such objects do as well and so a further excess of detections, beyond that identified here for CDM, could be found, giving evidence for SIDM over CDM. In this work we were not able to explicitly test this, as it would require sensitivity mapping the data over a range of concentrations. This is still a computationally intractable problem even with the speed of the machine learning method.

If the population effect can be properly accounted for in dark matter inferences from strong lensing, the constraining power of all available data dramatically increases. The Euclid telescope is already discovering the first of its $100,000$ new strong lenses \citep{Walmsley2025}. The space-based resolution of the data has been shown to be good enough to detect $\sim2500$ new subhaloes \citepalias{ORiordan2023}. These individual detections would be at the high end of the subhalo mass function $(\Mmax>10^{8.8}\Msun)$, which are useful for constraining the overall fraction of mass in substructure, but constraints on the dark matter particle properties are weak. If, as this work suggests, these data are sensitive to the collective effect of masses in a more strongly constraining range in terms of dark matter properties, the Euclid dataset becomes a much more powerful probe. This would also be the case for similar datasets from e.g. Roman or the Square Kilometre Array (SKA).
	\section{Conclusions}
\label{sec:conclusions}

In this paper we examined the effect of large populations of low-mass haloes on strongly lensed images. We showed that small subhaloes, themselves undetectable individually, have a detectable effect on the lensed images when in large numbers, which is the case in CDM. We summarise the method and the main results as follows.

\begin{itemize}
  \setlength{\itemsep}{3pt}      
  \setlength{\parskip}{0pt}        
  \setlength{\parsep}{0pt}
  \setlength{\topsep}{0pt}         
  \setlength{\leftmargin}{1em}    
  \setlength{\itemindent}{9pt}     
  \setlength{\labelsep}{0.5em}    
  \renewcommand\labelitemi{\small$\bullet$}
    \item We used a machine learning substructure detection model, trained on realistic data, to produce sensitivity maps for 100 simulated strong lens observations, of a similar quality and configuration to HST SLACS and BELLS GALLERY samples.
    \item The sensitivity map gives the smallest mass of subhalo detectable at $5\sigma$ in each pixel. Using this lower limit, we use a subhalo mass function, dependent on the dark matter model, to predict the number of detections in each lens for an average population of CDM subhaloes.
    \item We then produce simulated images of the same $100$ strong lenses containing exactly these CDM populations and pass them through the subhalo detector. We repeat this for many realisations of the subhalo population.
    \item If the only detectable subhalo signal is from those individual objects above the sensitivity threshold, then the sensitivity map should accurately predict the frequency of detections made by the model.
    \item We find a $40$ per cent excess in the number of detections in CDM, i.e., our subhalo detector finds subhaloes with $1.4$ times the frequency predicted by the sensitivity map. This excess dissipaters in warmer dark matter models, where the population of subhaloes is suppressed at the low mass end.
    \item Using a mass function sharply truncated at the lower end, we repeat the experiment. We find that the excess in the number of detections is not present when only masses in the sensitivity range of the data are included. As lower masses are included, the excess begins to grow, even though none of these objects are individually detectable. At very low masses, the excess stops increasing, indicating that only subhaloes in the intermediate mass range cause this effect.
    \item We propose a psuedo-mass function which we use to boost the number of detectable objects. We fit to the measured frequency of detections to obtain the parameters of this mass function. We find that, in CDM, it mostly consists of masses two orders of magnitude below the individual detection threshold.
    \item Finally, we repeat our analyses for subhalo detection models which include varying amplitudes of circular multipole perturbations to the lens galaxy. We find that the multipoles present a strong degeneracy with the signal of this low-mass population, much more so than for individually detectable objects. However, this result would change if more physically motivated elliptical multipoles were used, which may be less degenerate with the subhalo signal in general.
\end{itemize}

Future dark matter studies with strong lensing have the capability to increase their constraining power by properly accounting for this effect, especially for studies using large sky surveys like Euclid, Roman, and SKA. However, as with individual detections of subhaloes, an understanding of the mass structure of the lens galaxy is critical for avoiding systematic biases and incorrect inferences.

	\section*{Acknowledgements}
	
	The author thanks Simona Vegetti for useful comments on the manuscript.
	
	\section*{Data Availability}
	The data used in this paper are available from the corresponding author on request.

	\bibliographystyle{mnras}
	\bibliography{bibliography}
	
	\bsp	
	\label{lastpage}
	
\end{document}